\documentclass{aa}
\usepackage{graphicx}
\usepackage{longtable,lscape}
\usepackage{hyperref}
   \hypersetup{
     colorlinks,
     citecolor=blue,
     linkcolor=blue,
    }
\usepackage{natbib,txfonts}
\begin{document}
   \title{Multiwavelength VLBI observations of Sagittarius\,A*
}

   \author{R.-S. Lu
          \inst{1,2,3}
          \and
               T.~P.~Krichbaum\inst{1}
      \and A.~Eckart\inst{2,1}
          \and S.~K\"onig\inst{2}
          \and D.~Kunneriath\inst{2,1}
          \and G.~Witzel\inst{2}
      \and A.~Witzel\inst{1}
      \and J.~A.~Zensus\inst{1,2}
          }

   \offprints{R.-S. Lu}

   \institute{
Max-Planck-Institut f\"ur Radioastronomie, Auf dem H\"ugel 69, D-53121 Bonn, Germany\\
              \email{rslu@mpifr-bonn.mpg.de}
          \and
   University of Cologne, Z{\" u}lpicher Str. 77, D-50937 K\"oln, Germany
          \and
   Shanghai Astronomical Observatory, Chinese Academy of Sciences, 80 Nandan Road, 200030 Shanghai, China}

   \date{Received 04 December 2009 / Accepted 15 September 2010 }


  \abstract
{The compact radio, NIR, and X-ray source Sagittarius\,A* (Sgr\,A*),
associated with the super massive black hole at the center of the
Galaxy, has been studied with Very Long Baseline Interferometry (VLBI)
observations performed on 10 consecutive days and at mm-wavelength.}
{Sgr\,A* varies in the radio through X-ray bands and occasionally
shows rapid flux density outbursts. We monitor Sgr\,A* with VLBI, aiming
at the detection of related structural variations on the submilliarcsecond
scale and variations of the flux density occurring after NIR-flares.}
{We observed Sgr\,A* with the Very Long Baseline Array (VLBA)
at 3 frequencies (22, 43, 86\,GHz) on 10 consecutive days in May 2007
during a global multiwaveband campaign. From this
we obtained accurate flux densities and sizes of the VLBI structure,
which is partially resolved at mm-wavelength.}
{The total VLBI flux density of Sgr\,A* varies from day to day. The variability is correlated
at the 3 observing frequencies with higher variability amplitudes
appearing at the higher frequencies. For the modulation indices,
we find 8.4\,\% at 22\,GHz, 9.3\,\% at 43\,GHz, and 15.5\,\% at 86\,GHz. The radio spectrum is inverted between 22 and 86\,GHz, suggesting inhomogeneous synchrotron self-absorption with a turnover frequency at or above 86\,GHz. The radio spectral index correlates with the flux density, which is harder (more inverted spectrum) when the source is brighter. The average source size (FWHM) does not appear to be variable over the 10-day observing interval.
However, we see a tendency for the sizes of the minor axis to increase with
increasing total flux, whereas the major axis remains constant.
Towards higher frequencies, the position angle of the elliptical Gaussian
increases, indicative of intrinsic structure, which begins to dominate
the scatter broadening. At cm-wavelength, the source size varies with wavelength as $\lambda^{2.12\pm0.12}$, which is interpreted as the result of interstellar scatter broadening. After removal of this
scatter broadening, the intrinsic source size varies as $\lambda^{1.4 ... 1.5}$. The VLBI closure phases at 22, 43, and 86\,GHz are zero within a few degrees, indicating a symmetric or point-like source structure. In the context of an expanding plasmon model, we obtain an upper limit of the expansion velocity of about 0.1\,c from the non-variable VLBI structure. This agrees with the velocity range derived from the radiation transport
modeling of the flares from the radio to NIR wavelengths.}
   {}

\maketitle
%
\section{Introduction}
It is now commonly assumed that most, if not all galaxies harbor
super-massive black holes (SMBH) at their centers \citep[e.g.][]{1998Natur.395A..14R,2004cbhg.symp....1K}. The nearest of these is the $\sim 4\times10^6$ M$_{\sun}$ SMBH at the center of our galaxy (see \citealp{2009IJMPD..18..889R} for a recent review). Thanks to its relative proximity at a distance of only $\sim 8$ kiloparsecs \citep{1993ARA&A..31..345R, 2005ApJ...620..744G,
2003ApJ...597L.121E, 2008ApJ...689.1044G, 2009ApJ...692.1075G,2009ApJ...700..137R},
high angular resolution VLBI observations of Sgr\,A* offer a
unique opportunity for testing the SMBH paradigm.

The observed frequency-dependent apparent source size of Sgr\,A* is commonly
interpreted as coming from the scatter broadening effect of the
intervening interstellar medium \citep[e.g.][]{2006ApJ...648L.127B, 2006JPhCS..54..328K}. The $\lambda^2$ dependence of the scattering effect has been driving VLBI observations of Sgr\,A* to
shorter and shorter wavelengths, towards vanishing image blurring.
In fact, mm-VLBI observations of Sgr\,A* at
43 and 86\,GHz suggest a break in the $\lambda^2$ dependence of the
scattering law. This implies that the intrinsic source structure
becomes visible and begins to dominate the scatter broadening
effect above $\nu \simeq 43$\,GHz \citep{1998A&A...335L.106K, 1998ApJ...508L..61L, 2001AJ....121.2610D, 2004Sci...304..704B, 2005Natur.438...62S, 2006ApJ...648L.127B, 2006JPhCS..54..328K}. The
recent detection of Sgr\,A* with VLBI at 1.3 mm at a fringe spacing of
$\sim$ 60 $\mu$as has pushed the limit to the size of the compact VLBI emission
down to  $\sim 4$ Schwarzschild radii (size $\sim 43$\,$\mu$as). This
is smaller than the theoretically expected size of the emission of an accretion disk
around a
$4\times10^6$ M$_{\sun}$ SMBH, assuming its non-rotation \citep{2008Natur.455...78D}.
At present it is unclear whether the compact emission seen by 1.3 mm-VLBI is related to the (relativistically aberrated) silhouette of the accretion disk emission around the BH, a hot spot or inhomogeneity in the accretion disk, a jet nozzle, or to something else \citep[][and references therein]{2000ApJ...528L..13F,2000A&A...362..113F,2006MNRAS.367..905B,
2009ApJ...701.1357B, 2007MNRAS.379..833H}.

Sgr\,A* is found to vary in the radio to X-ray regime with
its activity more pronounced (larger amplitudes, shorter timescales)
at shorter wavelengths \citep{2001Natur.413...45B, 2003Natur.425..934G, 2004ApJ...601L.159G, 2006A&A...450..535E, 2006A&A...455....1E}. The short variability timescales (down to minutes) and a possible quasi-periodicity of $\sim$ 17--30 mins suggests that the variability
originates in a very compact region, possibly located near the last
stable orbit close to the event horizon of the BH. The study of the
frequency dependence of the variability and the observed time lags between frequencies can basically be explained via the expansion of an orbiting
synchrotron self-absorbed emission component, which cools down and decays after an initial flare \citep{2006A&A...450..535E, 2008ApJ...682..373M, 2008ApJ...682..361Y, 2008A&A...492..337E,2009ApJ...706..348Y}.

However, direct detection of transient structure component(s),
which could be directly related to such flaring events, is still pending.
From the decay of the observed flux densities and the
time delays of the emission peaks among X-ray, NIR, and short millimeter wavelengths, one derives subrelativistic expansion speeds and a size of the emission region of only a few Schwarzschild radii \citep{2009A&A...500..935E}.
Millimeter-VLBI observations of Sgr\,A*, mainly performed within larger coordinated multiwavelength flux-monitoring campaigns, are therefore important, as they may allow one to detect and relate possible structural variability on AU-scales to the flux density activity observed at shorter wavelengths (sub-mm, NIR, X-ray).

Here we present new results from VLBI observations of Sgr\,A*, which
took place during a global multifrequency campaign in May 2007
\citep[see][for a description of this campaign]{2008A&A...479..625E}.
Sgr\,A* was observed with the VLBA in dual circular polarization on
10 consecutive days during May 14 - 25, 2007 at 22, 43, and 86 GHz.
In this paper we focus on the data and results from the total intensity data. The signature of polarization is less clear due to the low
linear polarization of Sgr\,A* and will be discussed in a future paper. Some preliminary results from these VLBI experiments were already
reported earlier \citep[e.g.,][]{2008JPhCS.131a2002E, 2010A&A...517A..46K, 2008JPhCS.131a2059L, 2010A&A...510A...3Z}.
The paper is organized as follows: observations and data analysis are described in Sect. \ref{sec:2}. The results are presented in Sect. \ref{sec:3}, followed by a discussion in Sect. \ref{sec:4}.
Summary and conclusions are given in Sect. \ref{sec:5}.


\section{\label{sec:2}Observations and data analysis}
Sgr\,A* was observed on 10 consecutive days on May 14 - 25
with the VLBA\footnote{The National Radio Astronomy
Observatory is a facility of the National Science Foundation
operated under cooperative agreement by Associated Universities,
Inc.} for a duration of 8 hrs per day, spending about one third
of the time on each of the three observing frequencies. Each VLBI
station recorded dual circular polarization at a recording rate of
512 Mbps (8 intermediate frequency (IF) channels, 16 MHz per IF,
and 2 bits per sample). The target source (Sgr\,A*) and the calibrators (VX Sgr, NRAO\,530, PKS\,1749+096, 3C\,279, 3C\,446) were observed in a frequency switch mode, cycling between 86, 43, and 22\,GHz in a duty cycle of $11-13$\,mins between all 3 frequencies. At 22\,GHz, the individual VLBI scans lasted 3\,min, and 4\,min at 43 \& 86\,GHz. The bright quasars
PKS\,1749+096, 3C\,279, 3C\,446 were observed as fringe
tracers for total intensity and cross-polarization
in the beginning and at the end of each VLBI experiment each day.
The quasar NRAO\,530 was observed more regularly, serving as a fringe tracer and to provide checks of the consistency of the amplitude
calibration. The SiO maser in VX\,Sgr (transitions $v$=1, $J$=1--0 and
$J$=2--1) was observed in interleaved, short VLBI scans of 1\,min duration,
and only at 43 and 86 GHz. These spectral line
measurements were made to complement the regular system temperature measurements and for an improved amplitude calibration
via their auto-correlation functions. All data were correlated at the VLBA correlator in Socorro, NM, USA with 1 s integration time.

\begin{figure*}
\centering
\includegraphics[angle=0,scale=0.5]{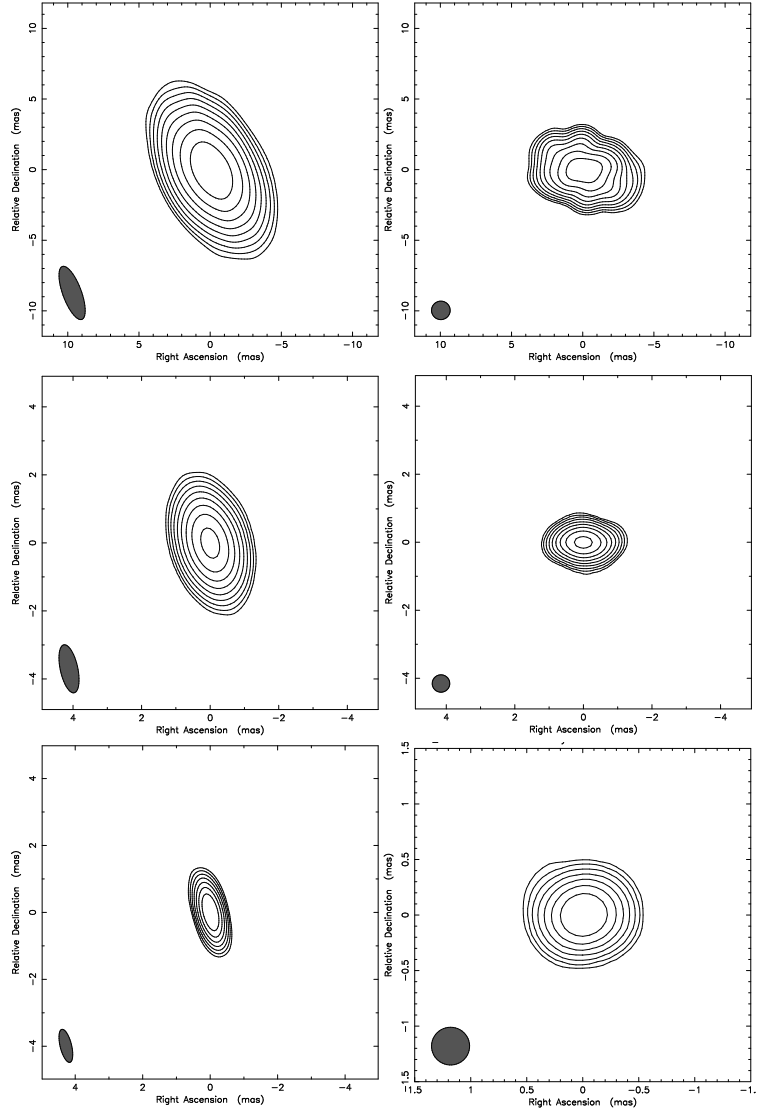}
\caption{Uniformly weighted VLBA images of Sgr\,A* on May 15, 2007  at 22 (top panels), 43 (middle panels), and 86\,GHz (bottom panels), respectively. At each row of panels, the righthand panel shows a slightly super-resolved image restored with a circular beam corresponding to the minor axis of the elliptical beam of the lefthand image. The contour levels are the same as in the lefthand panel. The parameters of the images are listed in Table~1.}
\label{fig:maps}
\end{figure*}

The data were analyzed in AIPS using the standard algorithms
including phase and delay calibration and fringe-fitting. The
amplitude calibration was performed using the measurements of the
antenna system temperatures and `a-priori' gain-elevation curves
for each station. Atmospheric opacity corrections were applied using the AIPS task ``APCAL'' from fits of the variation in the system temperature when plotted against air mass (so-called sky-dips).
Images of Sgr\,A* were finally produced using the standard hybrid mapping and CLEAN methods in AIPS and with DIFMAP at all three frequencies. In Figure \ref{fig:maps}, we show the resulting CLEAN
images at 22, 43, and 86 GHz as an example from the observation on May 15, 2007. For a better visualization, we also convolved
the maps with a circular restoring beam. The maps are shown in the right panels of Fig.~\ref{fig:maps}. In Table~\ref{tab:para} the parameters of these images are summarized. During the imaging process and the stepwise iterative amplitude self-calibration, the correctness of the station gain solutions was monitored and controlled via a comparison of gain solutions independently  obtained on the calibrator continuum sources NRAO\,530, PKS\,1749+096 and from the auto-correlations of the spectral line emission of VX\,Sgr.

\begin{table*}
\caption{\label{tab:para}Description of VLBA images of Sgr\,A* shown in Fig.~\ref{fig:maps}.}
\begin{center}
\begin{tabular}{@{}c*{15}{c}}
\hline
&&\multicolumn{3}{c}{Restoring Beam}\\
\cline{3-5}
Frequency&S$_{peak}$&Major&Minor&PA&Contours\\
\hline
\mbox{[GHz]}&[Jy/beam]&[mas]&[mas]&[deg]&\\
(1)&(2)&(3)&(4)&(5)&(6)\\
\hline
22&0.653 (0.456)&4&1.31&20.1&0.2, 0.4, 0.8, 1.6, 3.2, 6.4, 12.8, 25.6, 51.2\\
43&1.20 (1.08)&1.44&0.513&12.1&0.2, 0.4, 0.8, 1.6, 3.2, 6.4, 12.8, 25.6, 51.2\\
86&2.98 (2.89)&1.01&0.34&13.7&0.2, 0.4, 0.8, 1.6, 3.2, 6.4, 12.8, 25.6, 51.2\\
\hline
\end{tabular}
\end{center}
\scriptsize {Notes: (1) Observing frequency; (2) Peak flux density, the numbers in brackets correspond to the peak intensity of images in the right panel of each row in Fig.~\ref{fig:maps};
(3), (4), (5) Parameters of the restoring elliptical Gaussian beam: the full width at half maximum (FWHM) of the major and minor axes and the position angle (PA) of the major axis.
(6) Contour levels of the image, expressed as a percentage of the peak intensity.}
\end{table*}

VLBI observations at millimeter wavelengths suffer from a number of
limitations mainly caused by the more variable weather and higher
atmospheric opacity, and limitations of the telescopes (e.g. steeper
gain curves, larger residual pointing and focus errors),
which were mainly built for observations at the longer centimeter wavelengths. For telescopes in the northern hemisphere, the relatively low culmination height of Sgr\,A* at the VLBA sites requires a careful observing and calibration strategy. Despite our best efforts in this context, mm-VLBI observations of Sgr\,A* are still subject to residual calibration inaccuracies, which are larger than for sources of higher declination, and correspondingly higher antenna elevations. Owing to the low declination of Sgr\,A*, the uv-coverage is elliptical, resulting in a lower angular resolution in the north-south than in the east-west directions. This leads to an elliptical observing beam and corresponding lower positional accuracy for structural components, which may be oriented along the major beam axis
(see Table~\ref{tab:para}). However, since the source structure of Sgr\,A* is
very symmetric and almost point-like (zero closure phase, see Sect. \ref{sec:41}), the source size can be reliably estimated by fitting Gaussian components to the visibilities and by comparing that size with the size and orientation of the actual observing beam.

At 22 and 43 GHz, we imaged and self-calibrated the data of Sgr\,A* following
standard procedures. The overall antenna gain corrections are verified by those derived from self-calibration of relatively nearby sources, such as
NRAO\,530 and PKS\,1749+096. The agreement between corresponding gain
correction factors was found to be better than 10\,\%, probably thanks to relatively good weather conditions during the experiments. At these two frequencies, the total flux densities of NRAO\,530 are reproduced well, with rms-fluctuations between epochs at levels of 1\,\% and 2\,\%, respectively.
These numbers reflect the normalization uncertainty of the flux densities and are based on the assumption that the total flux density of NRAO\,530 is not variable at all 10 epochs. PKS\,1749+096 reproduces its flux densities slightly worse, mainly owing to the more limited uv-coverage. At 22 and 43\,GHz, we determine the overall accuracy of the
amplitude calibration to be $\sim 3-5$\,\%.

At 86\,GHz, the flux densities of Sgr\,A* were mainly determined on the basis
of the short uv-spacing data (between 20 and 100 Mega-lambda) and after
careful data editing. Owing to its lower elevation, the amplitude scaling factors applied to Sgr\,A* are in general somewhat larger than those applied to NRAO\,530, but they show the same overall trend within each experiment and over the 10 observing days. The observed deviations are in the range of the expected atmospheric opacity increase, since Sgr\,A* is $\sim$ 16$^{\circ}$ further south in declination than NRAO\,530. They also incorporate
weather-dependent residual gain errors, which appear from the time interpolation of gain solutions using nearby calibrator scans. The amplitude correction factors obtained from Sgr\,A* and NRAO\,530 agree within $\simeq 20$\,\%, resulting in an upper limit for the calibration scaling error of the absolute total flux density, which is consistent with the scatter of the visibilities within 100 Mega-lambda at 86\,GHz. We therefore conclude that the accuracy of the overall amplitude calibration at 86\,GHz is $\simeq 20$\,\%. The flux densities of NRAO\,530, however, are reproduced at a level of $\sim 6$\,\% (see also Table~\ref{tab:flux} and Sect.~\ref{sec:3.1}), which defines the repeatability of the amplitude
calibration on the 10 consecutive days at this level and allows the investigation of flux density variations of Sgr\,A* at 86\,GHz (Sect.~\ref{sec:3.1}). Some more details of the amplitude calibration
procedure are explained in Appendix~\ref{appendix}.

\begin{figure}
\centering
\includegraphics[angle=0,width=0.48\textwidth]{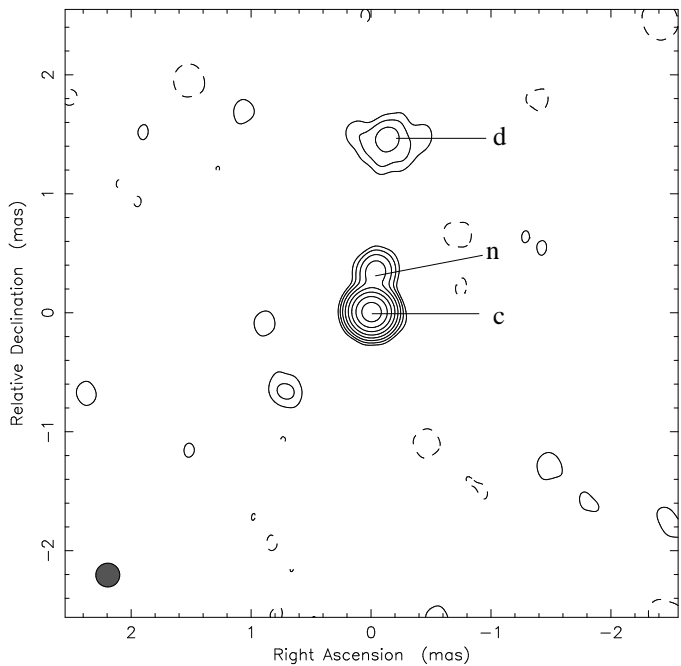}
\caption{A CLEAN image of NRAO\,530 at 86\,GHz on May 20, 2007. Contour levels are -0.5, 0.5, 1, ..., 64\,\% of the peak intensity of 1.13\,Jy/beam. The map was restored with a circular Gaussian beam of 0.2\,mas.}
\label{fig:nmap}
\end{figure}

The main secondary calibrator NRAO\,530 is core-dominated with the core dominance factor R ($\frac{S_{core}}{S_{total}}$) increasing with frequency. We show in Fig.~\ref{fig:nmap} a CLEAN map at 86\,GHz. At this frequency, R is $\sim$ 90\,\%, almost indicating a point source. The variation in the flux density ratio between core and jet (at each of the the three frequencies) also measures the effect of residual calibration errors, e.g., residual opacity effects. Using the flux density of the most prominent secondary jet component {\sl d} (see Fig.~\ref{fig:nmap}), which is visible at all three frequencies, the scale-free variations in the visibilities are found to be 1.5\,\% at 22\,GHz, 1.6\,\% at 43\,GHz, and 2.5\,\% at 86\,GHz. These numbers are consistent with the aforementioned repeatability errors at each frequency.

For the study of the structural variability, the VLBI source structure was parameterized in the usual way using Gaussian modelfits to the visibilities. Formal errors for the individual fit parameters were determined using the approach described in \citet{1998A&A...329..873K}. Briefly, as a measure of the uncertainty for each model parameter, we took the scatter from ``best fitting'' models, which were obtained from slightly different calibrated and edited data sets. These error estimates are more conservative than error estimates derived solely on the basis of rms noise in the map and SNR.
A more detailed description of the error analysis of the individual structure parameters is given in the corresponding sections of this paper.

\section{\label{sec:3}Results}
\subsection{\label{sec:3.1}Flux density variations and the spectrum}

\begin{table*}[ht!]
\caption{\label{tab:model} Results from modeling the data.}
\begin{center}
\begin{tabular}{*{8}{l}}
\hline\hline Date&$\nu$&S&$\theta_{major}$&Ratio&PA\cr &[GHz] &
[Jy]&[mas]&&[deg]\cr \hline May
15&22.2&1.51$\pm$0.02&2.56$\pm$0.03&0.59$\pm$0.04&78.6$\pm$1.7 \cr
      &43.1&2.02$\pm$0.09&0.71$\pm$0.01&0.58$\pm$0.06&81.7$\pm$2.3 \cr
      &86.2&4.06$\pm$0.57&0.25$\pm$0.02&0.68$\pm$0.08&79.8$\pm$8.0 \cr
May 16&22.2&1.18$\pm$0.02&2.55$\pm$0.06&0.50$\pm$0.03&82.0$\pm$1.1 \cr
      &43.1&1.59$\pm$0.07&0.72$\pm$0.01&0.53$\pm$0.04&82.1$\pm$0.8 \cr
      &86.2&2.89$\pm$0.83&0.19$\pm$0.01&1.00&- \cr
May 17&22.2&1.52$\pm$0.03&2.55$\pm0.04$&0.63$\pm$0.05&78.0$\pm$1.7\cr
      &43.1&1.99$\pm$0.06&0.72$\pm$0.01&0.62$\pm$0.05&82.0$\pm$2.4 \cr
      &86.2&3.96$\pm$0.90&0.21$\pm$0.01&0.81$\pm$0.16&82.5$\pm$6.7 \cr
May 18&22.2&1.23$\pm$0.02&2.56$\pm$0.05&0.60$\pm$0.03&79.8$\pm$1.0\cr
      &43.1&1.61$\pm$0.04&0.71$\pm$0.01&0.54$\pm$0.09&84.1$\pm$1.1 \cr
      &86.2&2.83$\pm$0.36&0.18$\pm$0.01&0.69$\pm$0.06&82.2$\pm$9.8 \cr
May 19&22.2&1.38$\pm$0.03&2.53$\pm$0.03&0.57$\pm$0.02&79.6$\pm$1.3\cr
      &43.1&1.86$\pm$0.07&0.71$\pm$0.01&0.51$\pm$0.06&84.7$\pm$3.1 \cr
      &86.2&3.68$\pm$0.83&0.19$\pm$0.02&1.00&-\cr
May 20&22.2&1.16$\pm$0.02&2.53$\pm$0.02&0.53$\pm$0.02&80.2$\pm$1.3\cr
      &43.1&1.66$\pm$0.06&0.72$\pm$0.01&0.54$\pm$0.06&80.4$\pm$1.6 \cr
      &86.2&3.24$\pm$0.37&0.20$\pm$0.02&0.54$\pm$0.13&86.4$\pm$4.4 \cr
May 21&22.2&1.42$\pm$0.04&2.58$\pm$0.02&0.61$\pm$0.02&77.2$\pm$0.8 \cr
      &43.1&2.02$\pm$0.08&0.72$\pm$0.01&0.62$\pm$0.06&81.8$\pm$3.1 \cr
      &86.2&4.18$\pm$0.50&0.23$\pm$0.01&0.69$\pm$0.08&80.4$\pm$7.7 \cr
May 22&22.2&1.36$\pm$0.02&2.56$\pm$0.03&0.58$\pm$0.02&78.9$\pm$0.5 \cr
      &43.1&1.90$\pm$0.05&0.72$\pm$0.01&0.66$\pm$0.06&78.5$\pm$1.3 \cr
      &86.2&3.73$\pm$0.73&0.21$\pm$0.01&0.54$\pm$0.09&84.4$\pm$11.8\cr
May 23&22.2&1.37$\pm$0.04&2.55$\pm$0.03&0.54$\pm$0.04&80.7$\pm$0.8\cr
      &43.1&1.92$\pm$0.08&0.72$\pm$0.01&0.54$\pm$0.06&81.2$\pm$1.3 \cr
      &86.2&3.26$\pm$0.57&0.18$\pm$0.02&0.62$\pm$0.10&77.7$\pm$6.1 \cr
May 24&22.2&1.35$\pm$0.01&2.57$\pm$0.01&0.54$\pm$0.02&81.1$\pm$0.9\cr
      &43.1&1.78$\pm$0.06&0.68$\pm$0.01&0.48$\pm$0.06&86.6$\pm$2.1 \cr
      &86.2&2.88$\pm$0.54&0.23$\pm$0.01&0.50$\pm$0.11&92.6$\pm$9.6 \cr
\hline
\end{tabular}
\end{center}
\scriptsize {Notes: Listed are the observing date in 2007, observing
frequency in [GHz], total flux density in [Jy], major axis of the
elliptical Gaussian in [mas], the ratio of the minor axis to the
major axis, and the position angle of the major axis.}
\end{table*}

\begin{figure*}
\centering
\includegraphics[angle=0,width=0.75\textwidth]{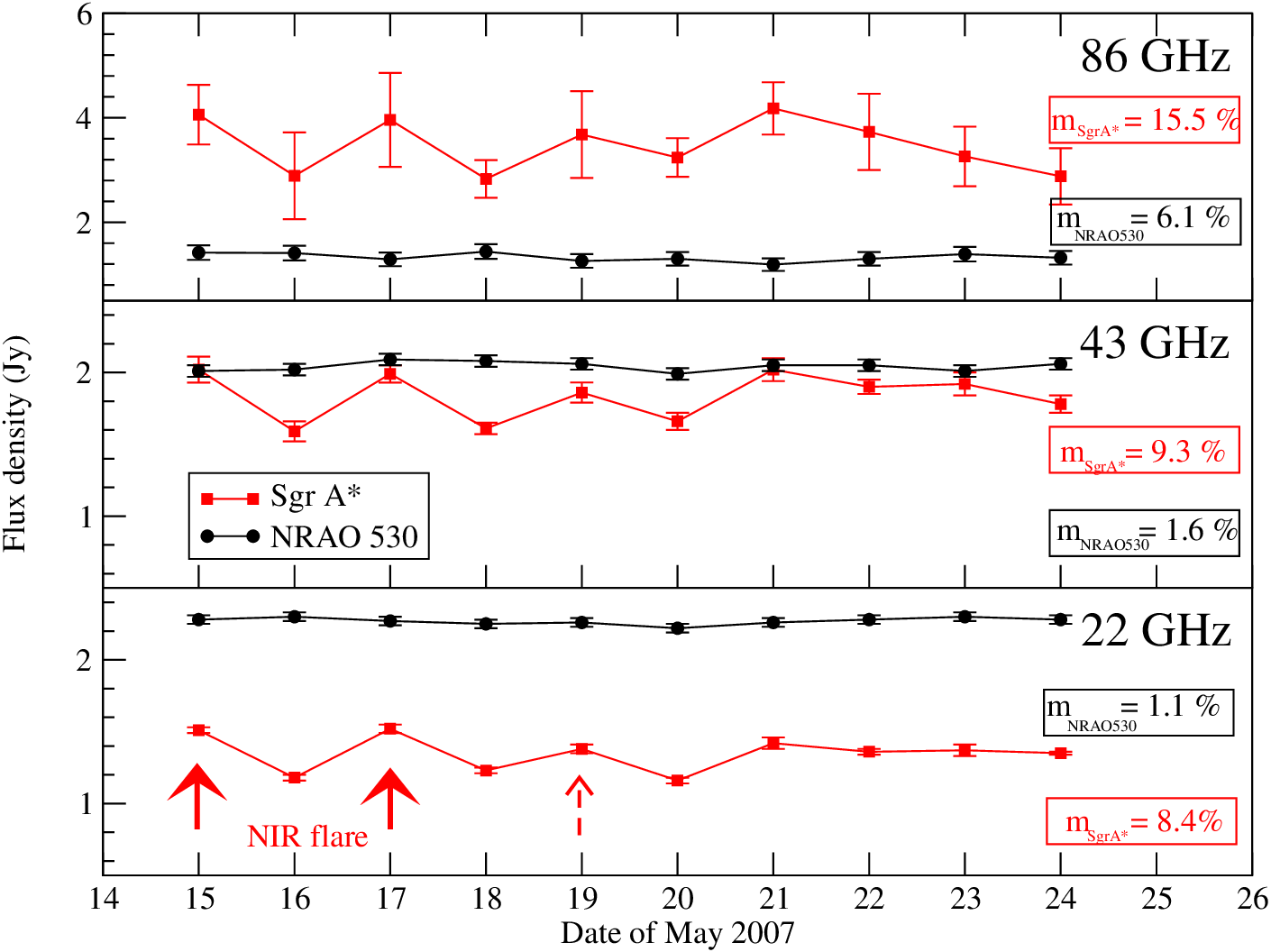}
\caption{Plot of flux density versus time for Sgr\,A* and nearby quasar NRAO\,530 at 22, 43, and 86\,GHz. Two filled arrows indicate times of NIR flares detected on May 15 \citep{2008A&A...479..625E} and 17. There is another possible mm flare (dashed arrow) on May 19 \citep{2010A&A...517A..46K}. In boxes, the variability index m is given for Sgr\,A* and NRAO\,530 at each frequency. The stationarity of the flux densities of NRAO\,530 characterize the quality and repeatability of the overall amplitude calibration.}
\label{fig:lc}
\end{figure*}

For Sgr\,A* we measured the total flux density by fitting a circular
Gaussian component to the edited and fully self-calibrated visibilities
of each VLBI observation after having made CLEAN maps for each epoch and using the Difmap software. The results of the individual model fits are shown in Table~\ref{tab:model}. Figure~\ref{fig:lc} shows the flux density of Sgr\,A* (and of NRAO\,530 used as a secondary calibrator) obtained from
each VLBI experiment on a daily basis at 22, 43, and 86\,GHz.
The 10 day average mean flux density is 1.33$\pm$ 0.04\,Jy at 22\,GHz, 1.79$\pm$0.05\,Jy at 43\,GHz, and 3.35$\pm$0.16\,Jy at 86\,GHz
(Table \ref{tab:mean}). The flux density variations of Sgr\,A* appear
more pronounced in the beginning of the campaign, during a time that coincides with two detected NIR flares occurring on May 15 and May 17, 2007
\citep{2008A&A...479..625E,2010A&A...517A..46K}. We defer the discussion of a possible relation of this variability with the variability at higher frequencies to Sect. \ref{sec:4}. The flux density variations of Sgr\,A* seen at 22, 43, and 86\,GHz appear highly correlated (indicated by the similar shape of the light curves) and progressively more pronounced towards the higher frequencies. We note that the measured flux densities of NRAO\,530 do not show such a pattern, reassuring us that the variations seen in Sgr\,A* do not come from calibration errors (see below).

\begin{table}[ht]
\caption{\label{tab:mean}The parameters of the time-averaged source model.}
\begin{center}
\begin{tabular}{*{7}{l}}
\hline\hline
$\nu$&S&$\theta_{major}$&$\theta_{minor}$&PA\cr
[GHz]&[Jy]&[mas]&[mas]&[deg]\cr \hline
22.2&1.33$\pm$0.04&2.56$\pm$0.01&1.44$\pm$0.03&79.5$\pm$0.4 \cr
43.1&1.79$\pm$0.05&0.71$\pm$0.01&0.40$\pm$0.01&82.0$\pm$0.6 \cr
86.2&3.35$\pm$0.16&0.21$\pm$0.01&0.13$\pm$0.01&83.2$\pm$1.5 \cr
\hline
\end{tabular}
\end{center}
\scriptsize {Notes: Listed are the observing frequency in [GHz], the
weighted mean of the total flux density in [Jy], the major axis of the
elliptical Gaussian in [mas], the minor axis in [mas], and of the
position angle of the major axis. Note that the minor axis and
position angles at 86\,GHz are averaged over all epochs where elliptical
Gaussian could be fitted.}
\end{table}

The average flux density at 22 GHz is comparable to previous measurements obtained during 1990-1993 \citep{2001ApJ...547L..29Z}. However, during our observations, Sgr\,A* appears brighter than in previous Very Large Array (VLA)
observations at 22 and 43\,GHz during 2000-2003 \citep{2004AJ....127.3399H}.
A `high' flux density state of Sgr\,A* is also seen at 86\,GHz,
where the present flux is much higher than previous flux densities
observed in 1996-2005 with the Nobeyama Millimeter Array (NMA)
\citep[mean value 1.1 $\pm$ 0.2\,Jy,][]{2005astro.ph.12625M}, and is comparable to the highest fluxes seen in October 2007 with the Australia Telescope Compact Array (ATCA) \citep[$\sim$ 4\,Jy,][]{2008JPhCS.131a2007L}.

In the following, we discuss the significance of a possible day-to-day
variability seen in the light curves of  Fig.~\ref{fig:lc}.
For comparison, we also plot the integrated total VLBI fluxes
of the nearby quasar NRAO\,530. In Table~\ref{tab:flux} we compare the variability indices $m$ (defined as the ratio of standard deviation and mean;
$m= \sigma / <S>$) between Sgr\,A* and NRAO\,530 at
the 3 frequencies. In all cases, Sgr\,A* shows higher values of m, and therefore variations with larger amplitudes than NRAO\,530 (see  Fig.~\ref{fig:lc}).

\begin{table}[ht]
\caption{\label{tab:flux}Flux density variability characteristics of Sgr\,A*.}
\begin{center}
\begin{tabular}{ccccc|ccc}
\hline\hline
&\multicolumn{4}{c|}{Sgr\,A*}&\multicolumn{3}{c}{NRAO\,530}\cr
\cline{2-5}\cline{6-8}
$\nu$&m&$\chi_\nu^2$&p&Y&m&$\chi_\nu^2$&p\cr
\hline
[GHz]&[\%]&&[\%]&[\%]&[\%]&&[\%]\cr
\hline
22.2&8.4&32.2&$\ll$0.01&25.0&1.1&0.6&76.2\cr
43.1&9.3&7.5&$\ll$0.01&27.5&1.6&0.7&73.0\cr
86.2&15.5&0.9&48.9&42.7&6.1&0.4&94.7\cr
\hline
\end{tabular}
\end{center}
\scriptsize{Notes: Listed are the observing frequency in [GHz], the
modulation index m, reduced $\chi_\nu^2$, probability p for the flux density being constant,
for Sgr\,A* and NRAO\,530, respectively. The variability amplitude (Y) is
calculated for Sgr\,A* (see text).}
\end{table}

We performed Chi-Square-tests to characterize
the significance of the variability, following e.g., \citet{2003A&A...401..161K}.
For the reduced $\chi_\nu^2$ we obtain values of
32.2 at 22\,GHz, 7.5 at 43\,GHz, and 0.9 at 86\,GHz.
The corresponding probabilities for the source not being variable are far less
than 0.01\,\% at 22 and 43\,GHz. At 86\,GHz, however, the probability for Sgr\,A* not being variable is only 48.9\% due to the larger measurement errors (see Table~\ref{tab:flux}).
Although being formally insignificant, the variations at 86\,GHz appear to
correlate with the variations seen at the two lower frequencies.
To describe the strength of the variability, the modulation index m
and the variability amplitude Y (defined as 3 $\times$ $\sqrt{m^2-m_{0}^2}$, where m$_{0}$ is the modulation index of the calibrator NRAO\,530) are summarized in Table~\ref{tab:flux}, where Y corresponds to a 3 $\sigma$ variability amplitude, from which systematic variations $m_0$, which are still seen in the calibrator, are subtracted \citep{1987AJ.....94.1493H}. For Sgr\,A* those from systematic bias corrected Y-amplitudes range between 25\,\% and 43\,\%.

The observed day-to-day variations compare well with similar variations
seen by other authors at other times. Using the VLA, \citet{2006ApJ...650..189Y} found an increase in the flux density at a level of 4.5\,\% at 22\,GHz and 7\,\% at 43\,GHz on timescales of 1.5-2\,hr. At 94\,GHz, \citet{2008JPhCS.131a2007L} observed intraday variability (IDV) with amplitude variations of 22\,\% within 2 hrs in August 2006, which confirmed  previously reported IDV by \citep{2005ApJ...623L..25M}.

Since for Sgr\,A* each VLBI track lasted about 6-7 hours, we are not
able to detect flux density variations on shorter timescales than this.
A splitting of the VLBI coverage at shorter intervals, e.g. in two or three coverages of equal duration, does not allow measuring the total source flux with sufficient accuracy (main limitation: uv-coverage and lack of secondary
calibrator scans) and therefore prevents the significant detection of variability on timescales shorter than 6\,hrs. A nonstationary source, which would vary with a large amplitude during the time of the VLBI experiment, however, would cause significant image degradation, leading to a reduced dynamical range in the CLEAN maps and the appearance of side lobes. Since this is not observed, we can exclude variations that are much larger than our typical amplitude calibration errors of $10-20$\,\%, on timescales shorter than the duration of the VLBI experiments.

From the measured total flux densities, we calculated a 3-frequency
spectral index between 22 and 86\,GHz (defined as S$_{\nu} \propto \nu^{\alpha}$).
We obtained an inverted spectrum, with spectral indices ranging between
(0.44 $\pm$ 0.04) and (0.64 $\pm$ 0.05). For Sgr\,A*, a frequency
break in the spectrum was suggested between $\sim 20 - 100$\,GHz
\citep{1998ApJ...499..731F, 2003ApJ...586L..29Z,
2005ApJ...634L..49A}. Below this break frequency, the spectral slope is much
shallower (lower) than at higher frequencies, where the so-called
sub-mm excess causes an increase in the inverted spectral index
\citep{1997ApJ...490L..77S, 2006JPhCS..54..328K}.
Our observing frequencies are just in the transition region between
cm- and sub-mm range. Therefore the measured spectral indices
are slightly higher than previously reported spectra, resulting
from VLBI at cm-wavelengths. \citet{2005ApJ...634L..49A} made simultaneous
multi-wavelength observations of Sgr\,A* in 2003. They describe the
spectrum from short centimeter (3.6\,cm) to millimeter (0.89\,mm)
wavelengths by a power law of the form S $\propto \nu^{0.43}$.
\citet{1998ApJ...499..731F} measured a spectral index of 0.52 between 7\,mm and 2\,mm wavelength. The observed spectral indices reported here are fully consistent with these previous studies, and confirm the onset of a sub-mm excess over the more shallow power-law shape seen at cm-wavelength.

\begin{figure}
\centering
\includegraphics[width=0.485\textwidth]{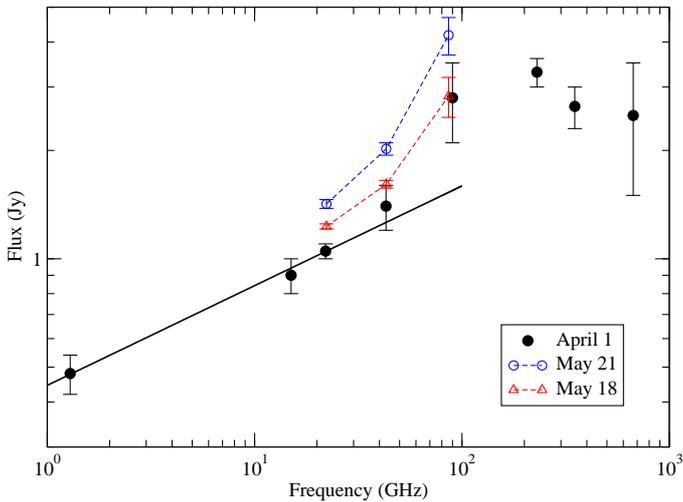}
\caption{Spectrum of Sgr\,*A. Filled circles denote a
quasi-simultaneous spectrum obtained around April 1, 2007 during a
multiwavelength campaign \citep{2009ApJ...706..348Y}.
The error bars on the data points indicate the
variability of Sgr\,A*. The solid black line indicates a power law fitted to
the radio data up to 43\,GHz. Above 43\,GHz, a flux density excess over this line is apparent. Open symbols connected by dashed lines indicate the new and quasi-simultaneous flux density measurements of May 18 and 21, 2007, from which high frequency spectral indices were derived, respectively. Note a spectral turnover between 100 -- 230\,GHz.}
\label{fig:spectruma}
\end{figure}

In Fig.~\ref{fig:spectruma}, we show spectra measured
at a ``quiescent'' (May 18) and during a ``flare'' (May 21) state.
A quasi-simultaneous spectrum from a multi-wavelength campaign taken
roughly one month before our observations is also shown for reference. Below $\sim$ 43\,GHz, the emission is characterized by an inverted spectrum. Towards higher frequencies,
the spectrum becomes more inverted with a spectral peak (turnover)
near 230\,GHz. For illustration of this ``excess'', we also show a power-law function fitted to the data below 43\,GHz, with a slope $\alpha \sim 0.3$. Both spectra from May 2007 show clear evidence of spectral curvature, in the sense that the 43 to 86\,GHz spectral index is more inverted than the spectral index between 22 and 43\,GHz. If we use the mean flux density (see Table~\ref{tab:mean}) obtained during our 10 days of observations, we find that the spectral index increases from 0.43 between 22 and 43\,GHz to 0.90 between 43 and 86\,GHz. This indicates that a new spectral component (the sub-mm excess) becomes clearly visible above $\nu \geq$ 43\,GHz.

If this high-peaking spectral component is interpreted as the result of synchrotron self-absorption (SSA) of a homogeneous synchrotron component, its magnetic field B can be calculated~\citep[see][]{1983ApJ...264..296M}:
\begin{equation}
B^{syn}=10^{-5}b(\alpha)\nu_{max}^5\theta^{4}S_{max}^{-2}\delta
\end{equation}
where $\nu_{max}$ is the peak frequency in GHz, $\theta$ the source size in milliarcsecond
using the FWHM for a Gaussian, $\theta_{G}$ ($\theta=1.8\theta_{G}$),
S$_{max}$ the peak flux density in Jansky, and b($\alpha$)
a tabulated parameter dependent on the spectral index, $\alpha$.
Adopting from Fig.~\ref{fig:spectruma} $\nu_{max}$ = 230\,GHz and the measured S$_{max}$ = 2.4 Jy, and as size $\theta_{G}$ = 37 $\mu$as~\citep{2008Natur.455...78D}, we obtain a
magnetic field $B^{syn}$ = 79.1 Gauss. In this calculation we assumed an optical thin spectral index above $\nu_{max}$ of $\alpha$ ($S_\nu \propto \nu^{\alpha}$) = -0.75 \citep{2006MNRAS.367..905B} and neglected relativistic boosting effects (Doppler-factor
$\delta=1$). The synchrotron cooling timescale ($t_{syn}$) is $\sim 5\times10^5\nu_{9}^{-0.5}B^{-3/2}$, where $t_{syn}$ is in minutes,$\nu_{9}$ is frequency in GHz, and $B$ in Gauss \citep{2009A&A...500..935E}. For our case, $t_{syn}$ is $\sim$ 50\,mins, which is consistent with the mm/sub-mm variability on hourly timescales.

\begin{figure}
\centering
\includegraphics[width=0.485\textwidth]{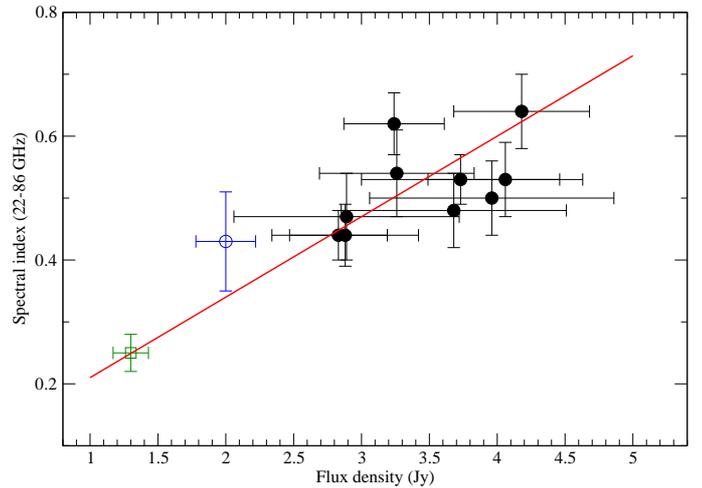}
\caption{Spectral index $\alpha$ (defined as S$_{\nu} \propto
\nu^{\alpha}$) as a function of flux density at 86\,GHz. Filled black
circles are data from this paper. Two additional data points at lower flux density levels
observed in earlier experiments
are added: \citet[][open circle]{1998ApJ...499..731F} and
\citep[][open square]{1997ApJ...490L..77S}. The solid line denotes the best
linear fit. See text for details.}
\label{fig:spectrum}
\end{figure}

Following \citet{2004AJ....127.3399H}, we plot the spectral indices as a
function of flux density at 86\,GHz in Fig.~\ref{fig:spectrum}. Our new data indicate that the 86\,GHz flux ($S_{\rm 86\,GHz}$) and spectral index ($\alpha$) may be positively correlated. The significance of this correlation is limited by the accuracy of the individual flux density measurements and by the limited range of flux densities ($S_{\rm 86\,GHz} \sim$ 2.8--4.2\,Jy) during our observations. We note, however, that spectral indices obtained earlier and in lower flux density states by \citet{1997ApJ...490L..77S} and \citet{1998ApJ...499..731F} confirm the trend seen in our new data, see the additional data points added to Fig.~\ref{fig:spectrum}\footnote{The original measurements of \citet{1998ApJ...499..731F} were made at 95\,GHz and are scaled to 86\,GHz, adopting a mean spectral index of 0.52}. By performing a linear regression analysis to all data, we obtain a correlation described by the following linear relation:
\begin{equation}
\alpha=(0.08 \pm 0.06) + (0.13\pm 0.02) \times S_{ 86 GHz}.
\label{eqno1}
\end{equation}

A correlation between spectral index and flux density at 86\,GHz
strongly favors source intrinsic flux density variability, similar
to the effect seen in AGN, where the (radio) spectrum hardens, when the source
is flaring. It seems very unlikely that interstellar scattering effects
can account for this correlation \citep{2004AJ....127.3399H}. The fast variability of the source flux, the high brightness temperature, and the correlation of the spectral index with the total source flux support the idea of a nonthermal synchrotron self-absorbed emission process with a spectral turnover frequency $\nu_{m}$ at or above 86\,GHz. This is consistent with the results of \citet{2008A&A...492..337E}, who also show that the turnover frequency is above 100\,GHz.

We conclude that in Sgr\,A* the observed variations in the total flux density are intrinsic to the source with several characteristics. (i) During the first 6 days (May 15 - 20), Sgr\,A* appears to be more variable than from day 7 onwards. This indicates there are different phases of activity. (ii) The variations appear correlated between frequencies and are more pronounced at
the higher frequencies. (iii) The study of the spectral variability shows spectral hardening during high states. (iv) The variations in the total VLBI flux density correlate well with similar variations of the total flux density seen between 90-230\,GHz at three other radio telescopes and possibly also with NIR flares \citep{2010A&A...517A..46K}.

\subsection{\label{sec:3.2}Source size measurement and its possible variability}

\begin{figure*}
\centering
\includegraphics[width=0.75\textwidth,clip]{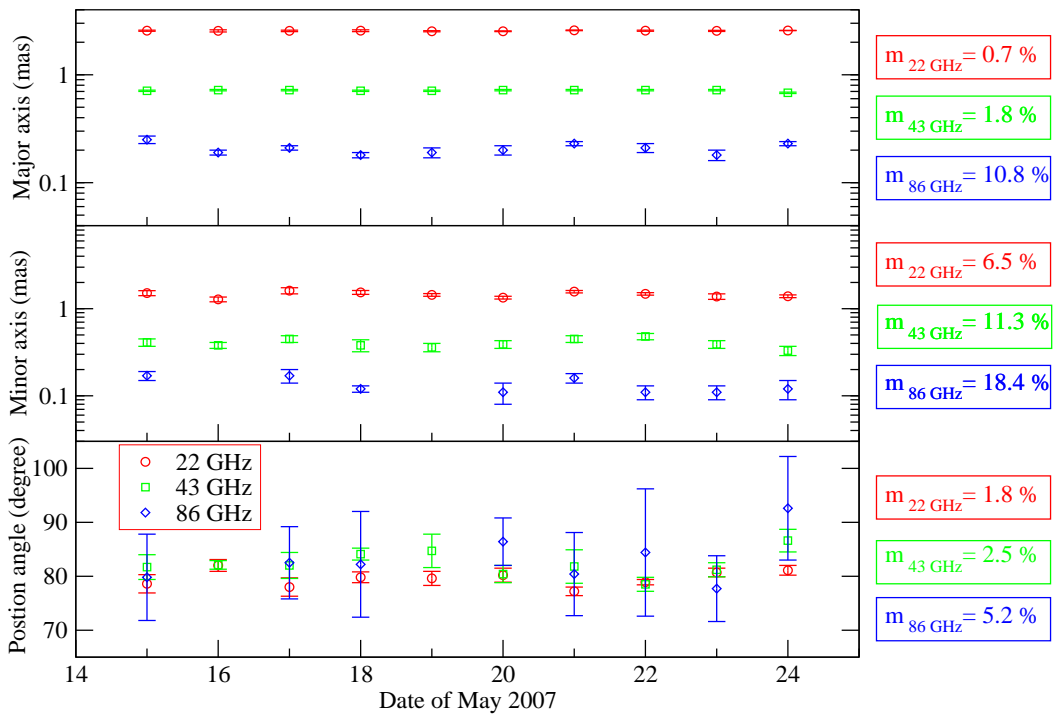}
\caption{Variability of the source structure of Sgr\,A* at 22 (circle), 43
(square), and 86\,GHz (diamond) between May 15 and 24, 2007. The
source structure is well-fitted by a elliptical Gaussian component. Shown
as a function of observing day are: the major axis (top
panel), the minor axis (middle panel), and the position angle of the
major axis (bottom panel). The corresponding variability indices m are
given in boxes on the right side. On May 16 and 19 of
May 2007, the data at 86\,GHz could only be fitted by
a circular Gaussian, therefore no minor axis and position angle
are given for these two dates.}
\label{fig:size}
\end{figure*}

At 22, 43, and 86\,GHz, the calibrated visibilities can be very well-fitted by a single elliptical Gaussian component. On May 16 and May 19, the elliptical fit to the 86\,GHz data diverged. In these two experiments we therefore fitted a circular Gaussian, and used the measured size for the size of the major axis. In Fig.~\ref{fig:size} we plot the major and minor axes and the position angle for the 10 individual VLBI experiments versus time. The top panel of the figure shows the major axis, the middle panel the minor axis, and the bottom panel the orientation (position angle) of the major axis. The different symbols denote the three different observing frequencies. In Table~\ref{tab:model} we summarize the results of the Gaussian model fitting. The observed mean values for sizes and position angles are also in good agreement with previous size measurements \citep{2004Sci...304..704B, 2005Natur.438...62S}. We note that some of these earlier results were obtained without fully calibrating the visibility amplitudes, but relied on closure amplitudes for the model fits.

The weighted mean and standard deviation of the time-averaged source parameters are summarized in Table~\ref{tab:mean}. The size of the major axis formally varies by rms/mean of m = 0.7\,\% at 22\,GHz, 1.8\,\% at 43\,GHz, and 10.8\,\% at 86\,GHz. Correspondingly, the minor axis changes by 6.5\,\% at 22\,GHz, 11.3\,\% at 43\,GHz, and 18.4\,\% at 86\,GHz.
The position angles at the three frequencies are all consistent with an east-west oriented source structure aligned along a position angle of $\sim$ 80.4 $\pm$ 0.9$^{\circ}$ (see Table ~\ref{tab:structure}).

\begin{table*}[ht]
\caption{\label{tab:structure}Structural variability characteristics of
Sgr\,A*.}
\begin{center}
\begin{tabular}{cccc|ccc|ccc}
\hline\hline
&\multicolumn{3}{c|}{Major axis}&\multicolumn{3}{c|}{Minor axis}&\multicolumn{3}{c}{PA}\cr
\cline{2-4}\cline{5-7}\cline{8-10}
$\nu$&m&$\chi_\nu^2$&p&m&$\chi_\nu^2$&p&m&$\chi_\nu^2$&p\cr
\hline
[GHz]&[\%]&&[\%]&[\%]&&[\%]&[\%]&&[\%]\cr
\hline
22.2&0.7&0.6&79.5&6.5&2.3&1.5&1.8&2.4&1.0\cr
43.1&1.8&1.6&11.9&11.3&1.3&21.8&2.5&2.0&3.7\cr
86.2&10.8&3.1&0.9&18.4&1.6&11.5&5.2&0.4&91.6\cr
\hline
\end{tabular}
\end{center}
\scriptsize {Notes: Listed are the observing frequency in [GHz], the
modulation index, reduced $\chi_\nu^2$, probability for the observable being constant for the major axis, minor axis, and position angle of the major axis, respectively.}
\end{table*}

For the major axis, our data indicate a formal non-variability of the source size at 22 and 43\,GHz. A formal $\chi^2$ test gives a 79.5\,\% probability that the major axis
at 22\,GHz is constant. At 43\,GHz we obtained an 11.9\,\% probability for a
constant size. However, at 86\,GHz we detect some marginal variability with a 0.09\,\% probability for constant value of the major axis (with m $\sim 11$\,\%). For comparison, we also measured the size of the VLBI core component of NRAO\,530 at 86\,GHz. Owing to the less dense uv-coverage and a more complex (core-jet) source structure, a fit of a single circular Gaussian component yields slightly larger variations of m $\sim 17.9$\,\%, which has to be corrected by a factor of $\sim \sqrt[]{\frac{N_{scan}(Sgr\,A*)}{N_{scan}(NRAO\,530)}}$ $\approx$ 2. The corrected fractional rms variation of NRAO\,530 is slightly less than the fractional variability obtained for Sgr\,A*, and is an upper limit to the systematic errors in the repeatability of the size measurements for Sgr\,A*, which then would be less than 9\,\%. We therefore conclude that, within the accuracy of our measurements, we did not detect any significant variations in the source size of the source structure at 22 and 43\,GHz. At 86\,GHz, however, the situation is less clear. The lower SNR of the visibilities and the larger uncertainties of the amplitude calibration, since they are affected more by low source elevation and higher and variable atmospheric opacities, make it difficult to judge the significance of the larger scatter seen at this frequency.

\begin{figure}
\centering
\includegraphics[width=0.485\textwidth]{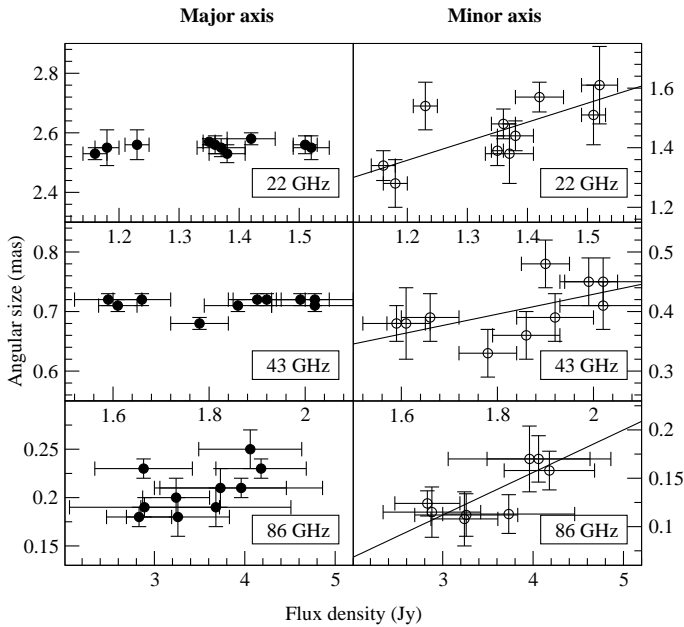}
\caption{Measured angular size as a function of flux density at 22 (top), 43 (middle),  and 86\,GHz (bottom). Both the major axis (left) and minor (right) axis sizes are shown. The solid lines delineate
best linear fits.}
\label{fig:size_flux}
\end{figure}

The existence of flux variability at millimeter wavelengths
naturally leads to the question of whether or not flux density variations
are associated with changes in size. As we have already noticed, the light curves (Fig.~\ref{fig:lc}) show strong variability at all three frequencies.
We now show in Fig.~\ref{fig:size_flux} the measured source size vs. flux density at each frequency, along with a linear fit. For the major axis size versus flux relation at 86\,GHz, the Spearman rank correlation coefficient, r$_{S}$, is 0.59. The probability of getting this by chance is less than 10\,\%, giving strong evidence for a correlation. For the minor axis,
a similar behavior is visible at all three frequencies. At 22\,GHz, there is a 3\,\% probability for random data leading to $r_{{\rm S}}$=0.72. The probability at 43\,GHz for the correlation arising by chance is 5\,\% for r$_{S}$=0.64. At 86\,GHz, r$_{S}$ is 0.54, with a 16\,\% probability of rejecting the non-correlation hypothesis.

The correlated variations in source size with flux density (for minor axis at all 3 frequencies, for the major axis at least at 86 GHz) suggest that they may be source intrinsic. The correlation seen along the minor axis could not be simply attributed to a sparse or variable uv-coverage along north-south directions because it is roughly the same across all 10 observing epochs at each frequency. In addition, for interstellar scattering the minimum timescale for the scattered size to change is roughly 10 months at 22\,GHz, 3 months at 43\,GHz, and 3 weeks at 86\,GHz \citep{2004Sci...304..704B}. These timescales are all larger than the duration of our 10-day observing campaign. In this context one may interpret the observed variations in the size of the minor axis as a consequence from a blending effect between interstellar scattering and source intrinsic size variation.

\begin{figure}
\centering
\includegraphics[width=0.485\textwidth]{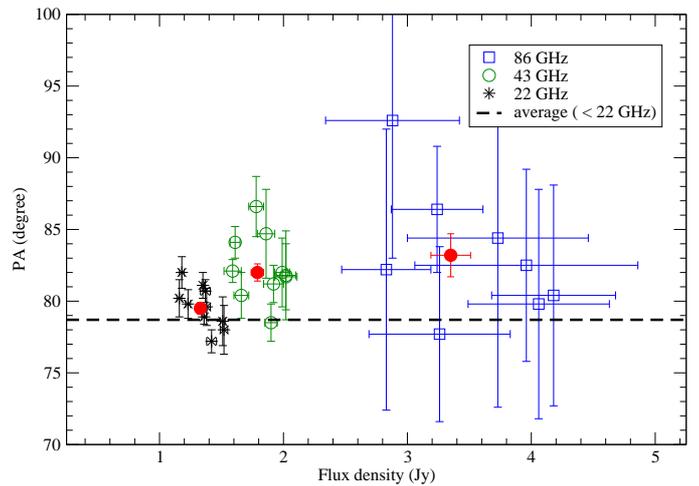}
\caption{Position angle of the major axis of Sgr\,A* plotted as a function of flux density for the 10-day VLBI observations at 22 (star), 43 (open circle), and 86\,GHz (open square). Also shown are weighted mean average values at each frequency (filled circles). The dashed line represents the mean position angle (78.8$\pm$1.7$^\circ$) at frequencies below 22\,GHz taken from literature data.
The position angles at 43 and 86\,GHz are significantly larger than the position
angle at 22 GHz and the lower frequency.}
\label{fig:pa_flux}
\end{figure}

For the position angle, we obtained 1.0\,\%, 3.7\,\%, and 91.6\,\% probability
for the source being constant at 22, 43, and 86\,GHz from formal $\chi^2$ tests (Table~\ref{tab:structure}). In Fig.~\ref{fig:pa_flux}, we plot the position angle of the major axis versus flux density for our 10 days of data. Because of the inverted spectrum at our observing frequencies, the data separate into 3 distinct point clouds. Their center of gravity is determined by a weighted average at each frequency. Compared with the average position angle obtained from literature data and taken at lower frequencies ($<$ 22\,GHz), the position angles show a tendency to increase towards higher frequencies (and because of the inverted spectrum: higher flux densities at higher frequencies). While the position angle at 22\,GHz is still consistent with values obtained at lower frequencies, the position angle at 43\,GHz deviates by $\sim 4.2\sigma$ from that at lower frequencies (see Table \ref{tab:mean}).
The position angle at 86\,GHz is consistent with the value at 43\,GHz, and is also marginally above the value at 22\,GHz. If confirmed by more accurate future observations, a possible interpretation for an increase of the position angle with frequency would be that scattering disk and intrinsic source structure are slightly misaligned and that the intrinsic structure of Sgr\,A* begins to dominate the east-west oriented scattering disk at the higher frequencies.

\subsection{Variations in the source size}
Recently, it has been claimed that Sgr\,A* shows variations in its size.
\citet{2004Sci...304..704B} claim an increase in its intrinsic size of $\sim$ 60\,\% in 2001 at 43\,GHz. \citet{2006JPhCS..54..377S} reported size variations at 43\,GHz for observations taken in 1999, where an increment of 25\,\% in its intrinsic size along the major axis has been detected. Eighty-six\,GHz VLBI observations performed in October 2005 between Effelsberg and IRAM suggest a size increase, which correlates with an increase in total flux density
\citep{2006JPhCS..54..328K}.

\begin{figure}
\centering
\includegraphics[width=0.485\textwidth,clip]{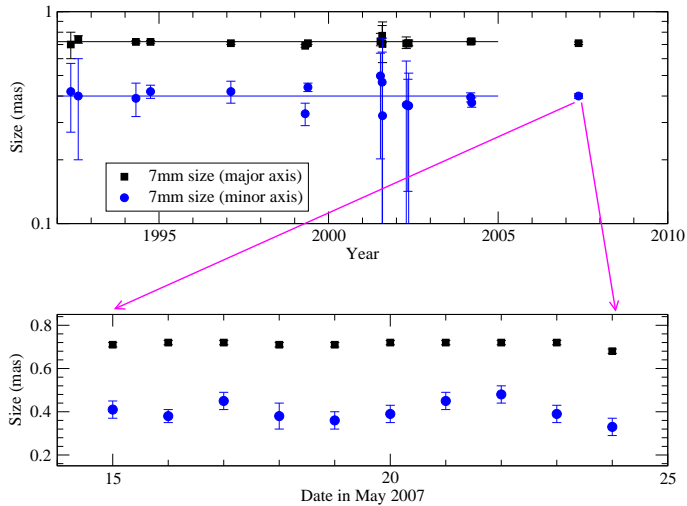}
\caption{Size of major axis (filled squares) and minor axis (filled circles)
plotted versus time at 43\,GHz during 1992--2007 (top panel). The solid lines delineate
an average of all previous measurements before 2007. The bottom panel
shows an enlargement on the data obtained during the 10 days campaign in May 2007.}
\label{fig:size_var_7mm}
\end{figure}

To study this further, we plot the size of the major and minor axes at 43\,GHz versus time
for all available observing epochs since 1992 in Fig.~\ref{fig:size_var_7mm}. The mean value of these sizes taken from the literature is 0.72$\pm$0.01 and 0.40 $\pm$0.01\,mas for the major and minor axes, respectively. The variability indices for the major and minor axis are 0.8\,\% and 8.2\,\%, consistent with the data of the new experiments presented in this paper. The corresponding probabilities of being variable are 75\,\% and 23\,\%, with reduced $\chi_{\nu}^2$ of 17.1 and 9.9 for the major and minor axes, respectively.

\begin{figure}
\centering
\includegraphics[width=0.485\textwidth,clip]{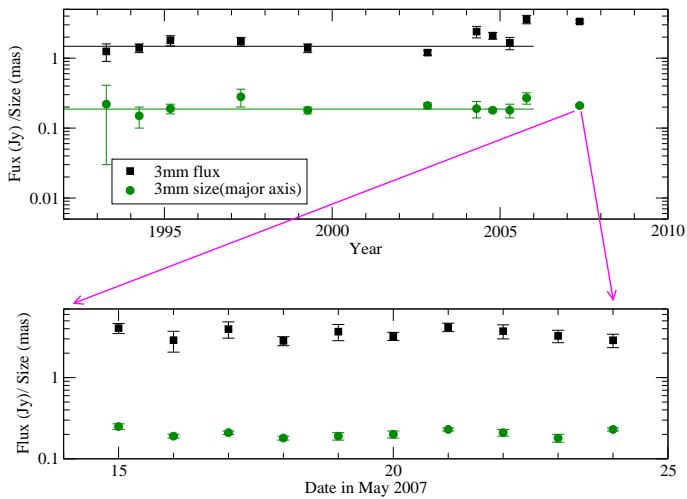}
\caption{Flux density (filled squares) and measured source size (filled circles) along the major axis of Sgr\,A* at 86\,GHz plotted versus time during 1993--2007 (top panel). The solid lines delineate the averages of all previous measurements before 2007. The bottom panel shows
an enlargement on the data obtained during the 10 days campaign in May 2007.}
\label{fig:size_var_3mm}
\end{figure}

In Fig.~\ref{fig:size_var_3mm} we also show the flux density and size of the major axis at 86\,GHz, comparing our new data with the data from the literature. The average flux density from the 10 days of VLBA data (3.35$\pm$0.16\,Jy) is close to the measurement obtained in 2005.79, and is about a factor of 2 higher than the mean value
of all previous flux density measurements (1.48$\pm$0.14\,Jy), indicating that the 86\,GHz flux density of Sgr\,A* varies on timescales of years. For the size, the average value from the ten VLBA observations is very close to the average value of all previous data (0.19$\pm$0.01\,mas), consistent with stationarity of the source size on timescales of a few
years. The limitations in the data quality (SNR), uv-coverage, and independent near in time measurements of the total flux density make it difficult to reliably establish or rule out a correlation between flux density and source size.
Although our data suggest mild variability of the source size at the 10-20\,\% level at 43 and 86\,GHz, the formal statistical analysis does not support significant variability of the size at any of the 3 observing frequencies. This prevents a more thorough correlation analysis between
VLBI sizes and the variability of the total flux density.

However we note that small amplitude variations on timescales of a few to less than one year appear to be present in the 86\,GHz data (Fig.~\ref{fig:size_var_3mm}).
Higher than average flux densities and sizes were recorded in 1997 and in 2005. It is clear that better VLBI data with higher SNR (larger observing bandwidth, larger telescopes)
and a better time sampling will be needed to study a possible correlation
between source size and flux density.

\subsection{Frequency dependence of source size}
\begin{figure}
\centering
\includegraphics[width=0.485\textwidth]{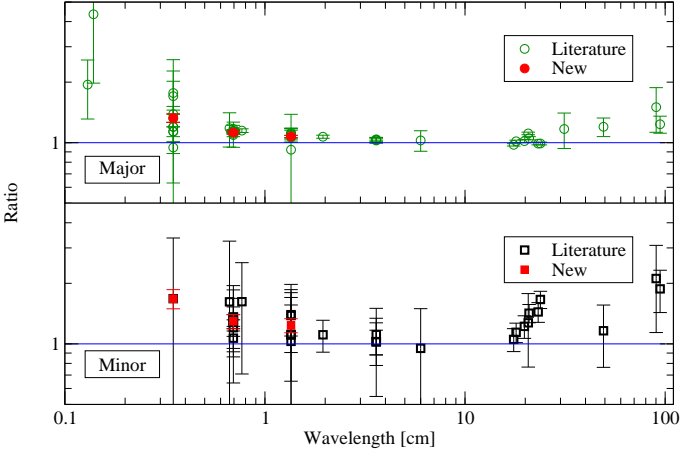}
\caption{The ratio between the apparent size of Sgr\,A* and the scattering size plotted as a function of wavelength for the major and minor axes. The scattering size is derived from the best fit of the scattering law in \citet{2006ApJ...648L.127B}, i.e., (1.309$\pm$0.015)$\lambda_{cm}^2$ mas and (0.64$^{+0.04}_{-0.05}$)$\lambda_{cm}^2$ mas for the major and minor axes, respectively. Data are from \citet{2005ApJ...634L..49A,2004ApJ...601L..51N,2003ANS...324..391R,1976MNRAS.177..319D,1994ApJ...434L..63Y,2006ApJ...648L.127B,2004Sci...304..704B,1989AJ.....98...44J,1985Natur.315..124L,1993Natur.362...38L, 1993AA...277L...1A,1999AA...343..801M,1993AA...274L..37K,1993Sci...262.1414B, 2005Natur.438...62S,1994cers.conf...39K, 1994ApJ...434L..59R, 1998A&A...335L.106K, 2006JPhCS..54..328K, 2001AJ....121.2610D,2008Natur.455...78D}.}
\label{fig:ratio}
\end{figure}

The intrinsic source size can normally be estimated
from the measured size by subtracting the scattering size in quadrature:
\begin{equation}
\theta_{int} = \sqrt{\theta_{meas}^2 - \theta_{scat}^2}.
\label{eqno2}
\end{equation}
The scattering deconvolved intrinsic size, therefore depends on the exact form
of the assumed scattering power law ($\theta_{scat} = a \times\lambda^{\zeta}$).
Recent revisions of this power law have assumed a $\lambda^{2}$ dependence of the scattering size \citep{2005Natur.438...62S,2006ApJ...648L.127B,2009A&A...496...77F}. Using our data and the data in the literature, we show in Fig.~\ref{fig:ratio} the ratio of the apparent major and minor
axes of Sgr\,A* relative to the $\lambda^2$-scattering model determined by \citet{2006ApJ...648L.127B}. Due to the poor constraint of the apparent size along the minor axis, it is difficult to see any difference in the broadening effect
between the major and minor axes. We notice, however, that both the major and minor sizes deviate from this scattering law at longer cm-wavelength. This discrepancy could stem from difficulties in measuring the size when faced with confusing extended emission around Sgr\,A* and in the Sgr\,A complex at long cm-wavelengths, but on the other hand, a steeper power-law index than 2 for the $\lambda$ dependence of the scattering size would remove this discrepancy completely. If this is the case, the intrinsic structure would begin to shine through already at somewhat longer wavelengths than currently estimated (e.g. at $\geq$ 3.6\,cm, \citet{2006ApJ...648L.127B}).

\begin{figure}
\centering
\includegraphics[width=0.485\textwidth]{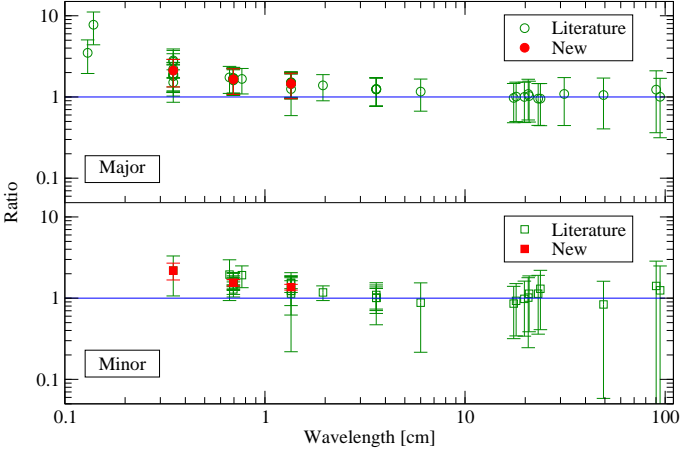}
\caption{Similar to Fig.~\ref{fig:ratio} except that the scattering size is derived from a slightly steeper power law of (0.93$\pm$0.32)$\lambda_{cm}^{2.12\pm0.12}$ mas for the major axis and (0.56$\pm$0.03)$\lambda_{cm}^{2.12\pm0.12}$ mas for the minor axis, respectively.}
\label{fig:ratio2}
\end{figure}

A power-law fit to the major axis size at wavelengths longer than 17 cm yields
for the wavelength dependence of the size the following power law:
\begin{equation}
\theta_{meas} = (0.93\pm0.32) \lambda_{cm}^{2.12\pm0.12} mas.
\label{eqno3}
\end{equation}
For this fit, a reduced $\chi_{\nu}^{2}$ of 18.9 is formally obtained, which
is lower than the reduced $\chi_{\nu}^{2}$ of 22.3 for the $\lambda^2$ scattering law proposed by \citet{2006ApJ...648L.127B}. In both cases, the
relatively higher value of the reduced $\chi_{\nu}^{2}$ indicates additional systematic
effects, which are not described well by the assumption of a simple power law.
We note that a  reduced $\chi_{\nu}^{2}$ of only 7 was obtained by \citet{2006ApJ...648L.127B},
when using a much narrower wavelength range of  17--24\,cm, i.e. when performing a local, not a global fit. For the minor axis, the sizes are determined less well. A direct power-law fit, which leaves both the slope and the coefficient unconstrained, does not yield a reasonable result. We therefore assume the same power-law index for the minor axis as for the major axis. With this restriction, we determine a normalization constant of 0.56$\pm$0.03 for the minor axis. Figure~\ref{fig:ratio2} shows the apparent size normalized by this scattering law. In comparison to the $\lambda^2$ dependence shown in Fig.~\ref{fig:ratio}, a systematic overshooting of the data over the model above 10\,cm wavelength is avoided.

The angular broadening of the size of Sgr\,A* is caused by the electron density fluctuations in the interstellar medium. The wavenumber (k) power spectrum of the density fluctuations is usually written in the form of a power law $\propto k^{\beta}$ with cutoffs on the largest (``outer scale", on which the fluctuations occur) and smallest (``inner scale", on which the fluctuations dissipate) spatial scales and power-law index $\beta$. The wavelength dependence of the scattering size follows $\theta_{scat} \varpropto \lambda^{\zeta}$, with $\zeta = \frac{\beta}{\beta-2}$ \citep[][and reference therein]{1998ApJ...508L..61L}. The angular broadening scales as $\lambda^{2.2}$ if the electron density spectrum is a power law with a Kolmogorov spectral index, $\beta = \frac{11}{3}$. When the length of the VLBI baseline becomes comparable to the inner scale, the scattering law changes and has the following form: $\theta_{scat} \varpropto \lambda^{2}$ \citep[e.g.,][]{2004ApJ...613.1023L}.

\citet{2004Sci...304..704B} shows that the source was exactly Gaussian in shape ($\beta=4.00\pm0.03$), indicating that the power-law index should be 2. They argue that the projected baselines are much shorter than the inner scale of scattering, and therefore $\zeta=2$ \citep{1989MNRAS.238..963N,1994MNRAS.269...67W}. The power-law index of $\zeta = 2.12 \pm 0.12$ determined at longer wavelengths (see above) corresponds to $\beta=3.8 \pm 0.2$, which is close to the Kolmogorov value of 3.67. Our findings therefore may indicate the presence of a broken power law, with a $\lambda^2$-dependence at intermediate wavelengths (cf. Fig.~\ref{fig:ratio}).
It was pointed out by \citet{1994MNRAS.269...67W} that the wavelength dependence of the
scattering size can show a break, depending on the size of the inner scale
for the scattering medium \citep[see also][]{2001ApJ...560..698L}. Towards the radio source
B1849$+$005, \citet{2004ApJ...613.1023L} finds a break in $\beta$ (their $\alpha$) between 0.33 and 5\,GHz with $\beta$ = 3.68 (i.e., $\zeta$ = 2.19) at 0.33\,GHz and $\beta$ $>$ 3.8 (i.e., $\zeta < 2.11$) above 5\,GHz. They interpret this break as evidence for detecting an inner scale of a few hundred kilometers in size, which is within the covered range of baselines from a few ten (VLA) to thousand (VLBA) kilometers. Applied to our data, we find $\zeta$ $>$ 2, if we fit the scattering size also including the longer wavelength (see Fig.~\ref{fig:ratio2}). If we only use data between 2 and 24\,cm, we obtain a slope $\zeta=1.99\pm0.01$, which is indistinguishable from 2.

\subsection{Intrinsic source size}

After subtracting the scattering law, we could determine the wavelength dependence of the source's intrinsic size. On the basis of the presently
available VLBI data, we fitted a power law ($\sim\lambda^{\gamma}$)
to the intrinsic source size in the wavelength range from
13 to 1.3\,mm. The intrinsic sizes were derived using Eq.~\ref{eqno2} and the two different versions of the scattering model.

\begin{figure*}
\centering
\includegraphics[scale=0.5]{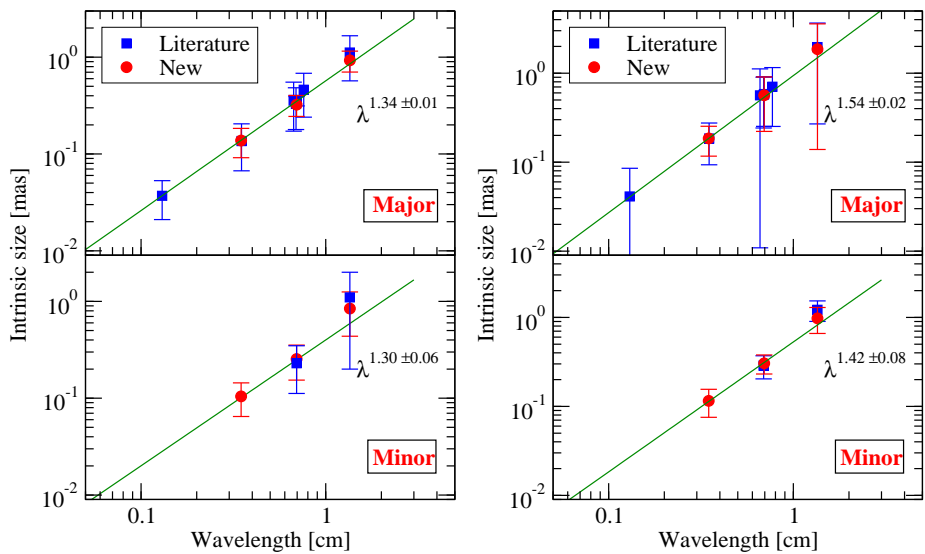}
\caption{Wavelength dependence of the source's intrinsic size for the major and minor axes. Left: Intrinsic size deconvolved with the $\lambda^2$ law from \citet{2006ApJ...648L.127B}. Right: The intrinsic size is derived from the steeper scattering law as used in Figure~\ref{fig:ratio2}. Literature data are taken from \citet{2004Sci...304..704B}, \citet{2005Natur.438...62S}, \citet{2006ApJ...648L.127B}, \citet{2008Natur.455...78D}.}
\label{fig:intrinsic}
\end{figure*}

Figure~\ref{fig:intrinsic} (left panels) shows fits assuming
a $\lambda^2$ scattering law, and Fig.~\ref{fig:intrinsic} (right panels) uses
$\lambda^{2.12}$ according to Fig.~\ref{fig:ratio2} and equation \ref{eqno3}.
While for the $\lambda^2$ dependence, the power-law indices of the major
and minor axes are 1.34$\pm$0.01, and 1.30$\pm$0.06, the steeper
scattering law, leads to a more inverted power-law index for the source intrinsic size, with 1.54$\pm$0.02 for the major axis and 1.42$\pm$0.08 for the minor axis. We note that \citet{2009A&A...496...77F} assume a $\lambda^2$ scattering law and use an index of 1.3$\pm$0.1 for the intrinsic source size, which is close to the slope obtained
above (see Fig.~\ref{fig:intrinsic}, top left).

\section{\label{sec:4}Discussion}
\subsection{\label{sec:41}Structural variability}
In the NIR, quasi-periodic flux density variations
have been claimed by several authors, e.g., \citet{2003Natur.425..934G},
\citet{2004A&A...417...71A}, and \citet{2006A&A...455....1E}.
The orbiting spot model is very successful
in describing these flare properties and their propagation through the spectrum. Supposing part of the emission comes from material orbiting around the SMBH, short-timescale
asymmetric structures would be expected \citep{2006MNRAS.367..905B}.
Thus quasi-periodic deviations of the closure phase from zero could be
expected on similar timescales, if the interferometer provides
a high enough angular resolution \citep{2008arXiv0807.2427F}. The closure
phase, the sum of three baseline phases in a closed triangle of
stations, is a phase quantity of the complex visibilities
that is independent of all antenna-based phase errors. The closure phase for any point-symmetric brightness distribution must be zero. In the context of the hot spot model, the degree and
timescale of a deviation of the closure phase from zero is a function of the hot spot orbital size, its inclination, and the flux density ratio between emission from the accretion
disk and the hot spot, embedded in it. Through phase-referenced VLBI monitoring
observations of the Sgr\,A* centroid position against stationary background quasars, \citet{2008ApJ...682.1041R} rule out hot spots with orbital periods that exceed 120 min and contribute $>$ 30\,\% of the total 7 mm flux density. However, structural variations at smaller radii or fainter hot spots still remain possible.

\begin{figure}
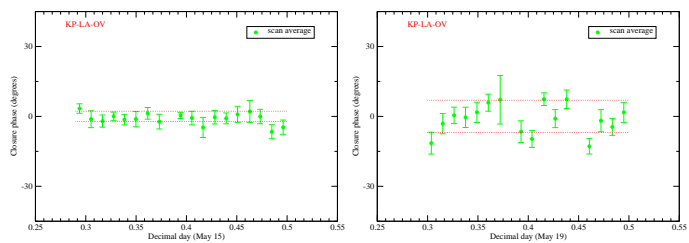

\centering
\includegraphics[width=0.24\textwidth,clip]{fig13.eps}
\includegraphics[width=0.24\textwidth,clip]{fig14.eps}
\caption{Plot of a typical closure phase as function of time for the KP-LA-OV triangle from the 86\,GHz experiment on May 15 (left) and 19 (right), 2007. In each plot, dotted lines indicate a
1 $\sigma$ range for the closure phase averaged over a whole experiment.}
\label{fig:cp}
\end{figure}

In an idealized and simplistic orbiting hot spot scenario, in which
the hot spot remains visible for several orbiting
periods, periodic deviations of the closure phase from zero could be expected
during the time of a VLBI track. This could be directly
observed with future mm/sub-mm VLBI arrays \citep{2009ApJ...695...59D}.
In this context, it is useful as a first step, to search for any
deviation in the closure phase from zero in the VLBI data presented here.
Owing to the smaller beam size and lower source intrinsic opacity,
such variations would be more pronounced and easier to detect at the
higher frequencies. We therefore used the closure phases at 86\,GHz for a more
detailed analysis. We extracted the closure phases from 10-second averaged $uv$-data for several representative station triangles, flagging discrepant phase data points before, and then coherently averaging the closure phases to the full scan length of about 4 minutes. In Fig.~\ref{fig:cp}, we show two typical examples of such closure triangles.

We now investigate the distribution of the closure phases and their errors in each triangle. Within a single VLBI run, the $\chi^{2}$-tests show that the closure phase remains constant with a high probability for the selected closure triangles. In a few cases where we found some marginal evidence of variability, a visual inspection of the data revealed a higher level of
phase noise and more outliers, which we could attribute to faulty data points. After their removal, we concluded that the closure phases do not vary systematically on a scan-by-scan basis (i.e. on timescales of a few minutes) and within a single VLBI run (timescale of hours). To further increase the signal-to-noise ratio and improve on the statistics, we therefore could average the closure phases in time. For each studied triangle, we obtained a characteristic mean closure phase values on any given day. In Table \ref{tab:clp3mm2} we summarize the results of our closure phase analysis and give averaged closure phase values and errors as a function of time for some representative baseline triangles of the VLBA. We use the standard deviation as the error of the mean. The three bottom lines of the table summarize the mean closure phase averaged for all observing epochs, the corresponding reduced $\chi^2_{\nu}$, and the probability for variability. The three right most columns of the table summarize the mean closure phase averaged over independent triangles at the given epoch, the reduced $\chi^2_{\nu}$, and the probability for variability. For all studied cases, the average closure phase is $0.1^\circ\pm1.6^\circ (1 \sigma)$, and for each individual triangle and observing day, the averaged closure phase is zero within its errors suggesting that the source structure is indeed point-like and not variable within a given day and on timescale of days.
This can be understood in terms of the 86\,GHz VLBI beam size of $\sim$ 0.3\,mas, which is a factor of $\sim$ 10--15 larger than the expected orbit size of a rotating hot spot, which must be larger than the size of the last stable orbit around the SMBH (1 Schwarzschild radius corresponds to about 0.01\,mas).

\begin{table*}
\caption{\label{tab:clp3mm2}Averaged closure phases for some representative
triangles at 86\,GHz.}
\begin{center}           
\begin{tabular}{c r r r r r r|r r c}        
\hline              
Date &KP-LA-OV  &FD-LA-PT    &FD-LA-OV    &FD-KP-PT    &LA-OV-PT    &FD-KP-OV&mean&$\chi_\nu^2$&Probability \\
\hline
(May 2007)&&&&&&&&(\%)\\
\hline                        
15 &$-0.5\pm$2.2&...        &...        &...        &$-1.2\pm3.1$&...&...&...&...\\
16&$-1.8\pm4.5$&$-3.2\pm5.9$&$-0.5\pm7.8$&$-1.4\pm6.9$& 0.0$\pm$8.2& 2.2$\pm$7.0&$-1.2\pm0.7$&0.08&0.5\\
17& 4.2$\pm$7.0&$-1.1\pm6.0$& 1.3$\pm$12.4&$-0.2\pm4.6$& 1.4$\pm$3.6&$-5.4\pm10.0$&0.6$\pm$0.9&0.16&2.2\\
18&$-1.4\pm4.0$& 2.0$\pm$5.3&$-2.1\pm4.2$& 0.7$\pm$6.8&$-2.7\pm5.6$& 3.4$\pm$9.6&$-0.8\pm0.8$&0.15&2.0\\
19&$-1.2\pm6.9$& 1.9$\pm$4.8& 2.8$\pm$7.3& 2.8$\pm$4.7&$-2.9\pm4.5$& 3.7$\pm$6.6&0.9$\pm$1.1&0.25&6.1\\
20& 2.6$\pm$4.9& 1.1$\pm$1.6&$-2.2\pm4.7$&$-0.2\pm4.0$&$-1.3\pm3.5$&$-4.2\pm8.5$&0.4$\pm$0.6&0.25&6.0\\
21& 4.1$\pm$6.5&$-0.5\pm2.8$&$-5.4\pm8.1$& 0.4$\pm$4.8&$-0.5\pm5.0$& 1.0$\pm$5.2&$-0.1\pm0.8$&0.19&3.2\\
22& 0.2$\pm$3.2& 0.9$\pm$2.0& 0.2$\pm$5.6&$-0.4\pm3.3$&$-1.3\pm4.3$&$-0.7\pm4.2$&0.2$\pm$0.3&0.06&0.3\\
23& 0.7$\pm$5.3& 0.5$\pm$3.4& 3.6$\pm$4.0& 1.8$\pm$5.8& 1.8$\pm$4.9& 1.8$\pm$6.9&1.7$\pm$0.5&0.08&0.4\\
24& ...        &$-1.1\pm3.2$&...         &$-1.1\pm4.1$&...         &...&...&...&...\\
\hline
mean &0.0$\pm$0.6&0.5$\pm$0.4&$-0.1\pm1.0$ &$0.1\pm0.4$&$-0.7\pm0.5$&0.5$\pm$0.9\\
$\chi^2_{\nu}$&0.18 &0.15&0.27&0.07&0.13&1.2\\
Probability(\%)&0.6&0.3&3.4&0.02&0.2&0.8\\
\hline                                 
\end{tabular}
\end{center}
\end{table*}

\subsection{\label{sec:42} Variability of VLBI source flux and NIR variability}

Sgr\,A* shows correlated flux density variability across the spectrum, from radio to NIR and X-ray. Early evidence of a broad-band relation comes from correlated flare activities of Sgr\,A* in the radio and X-ray bands \citep{2004ApJ...603L..85Z}. Recently, \citet{2008A&A...492..337E} and \citet{2009ApJ...706..348Y} have detected simultaneous flare emission in the near-infrared and sub-mm domains. \citet{2008A&A...492..337E} showed strong flare activity in the 0.87\,mm (345\,GHz) sub-mm wavelength band, following a NIR flare with a delay of 1.5$\pm$ 0.5 hours. Flares seen in the NIR and X-ray regime appear to happen almost synchronously \citep[see, e.g.,][]{2008A&A...479..625E}. Such high-energy flares apparently propagate down into the sub-mm/mm regime with a typical delay  of $\sim$ 100 minutes \citep[]{2008ApJ...688L..17M, 2008ApJ...682..373M, 2008ApJ...682..361Y}.

The NIR observations of Sgr\,A* on May 15, 2007 started at UT 5$^{\rm h}$29$^{\rm m}$ and lasted for 250 minutes. During that time, an NIR flare occurred at UT 7$^{\rm h}$30$^{\rm m}$ and peaked at UT 8$^{\rm h}$0$^{\rm m}$ \citep{2008A&A...479..625E}. The VLBI observations started between UT 6$^{\rm h}$10 - 6$^{\rm h}$20$^{\rm m}$ at the three frequencies and lasted for 6 hours. Therefore, the VLBI observations covered $\sim$ 4.2 hours (UT 8$^{\rm h}$0$^{\rm m}$ - UT 12$^{\rm h}$10$^{\rm m}$) after the peak time of the NIR flare. During this time we did not detect any significant change in the visibilities. Based on the assumption that the variations in the NIR flux density lead to a change in the source size \citep[see, e.g.,][]{2009A&A...500..935E}, we can impose limitations on the expanding speed of the putative expanding plasmon, so we use this 4.2 hour interval as an upper limit.
With the beam size of 0.34\,mas (0.1\,mas $\simeq$ 1 AU) at 86\,GHz,
the expansion speed would be limited to $<$ 0.1\,c. In other words,
any significant change in the source structure with speed greater than
this limit should have been detected. If, on the other hand, we use the 10-day interval as an upper limit, then we find an upper limit of 0.002\,c for the expansion speed. Modeling of radio, sub-mm, and NIR flares yields expansion velocities of 0.003--0.1 c \citep{2008ApJ...682..361Y, 2008A&A...492..337E}. This agrees well with the above expansion velocity, although the accuracy of the derived expansion speed from VLBI is still relatively poor.

\section{\label{sec:5}Conclusion}
We presented multi-epoch multi-frequency (22, 43, 86\,GHz) high-resolution
VLBA monitoring observations of the compact radio source Sgr\,A* in the Galactic center, aiming to search for structural variability on inter-day timescales. Sgr\,A* shows flux density variations on daily timescales with variability amplitudes of a few to a few ten percent. The variability amplitudes are systematically increasing with frequency. A positive correlation was found between the spectral index and the flux density at 86\,GHz (harder spectrum when brighter). The major axis size of Sgr\,A* appears to be stationary (nonvariable) at 22 and 43\,GHz.
Marginal variations seen at 86\,GHz are affected by residual measurement uncertainties and need more confirmation. In contrast to the apparent stationarity of the major axis, we found evidence of variability in the minor axis with time and flux density. This variability in the size of the minor axis points towards a source intrinsic origin. The source orientation (position angle of the major axis) appears to change with frequency. At 86 and 43\,GHz, the position angle is misaligned by about $2 \pm 1$ degrees, relative to the position angle measured at lower ($< 22$\,GHz) frequencies. This points towards a misalignment of the scattering disk and the source intrinsic structure, which begins to shine through towards high frequencies.

Our observations confirm the sub-mm excess in the radio spectrum of Sgr\,A* with the turnover frequency between $\sim$100--230\,GHz. The high-peaking spectral component (the sub-mm excess) could be due to synchrotron self-absorption (SSA) of a homogeneous synchrotron component. The estimated magnetic field strength of 79\,Gauss is consistent with hour timescale variability of a synchrotron source.

The commonly assumed $\lambda^2$-scattering law for the interstellar image broadening underestimates the observed angular size of Sgr\,A* at longer wavelengths for both the major axis and minor axis. A fit with a steeper power-law index of $2.12\pm0.12$ for the $\lambda$-dependence of the scattering size can remove this discrepancy. This may suggest that
the inner scale for the turbulence is a relevant parameter, which when taken into account makes the scattering towards the Galactic center similar to several other scatter broadenend radio sources. In the case of a steeper scattering law, the critical wavelength at which the intrinsic size begins to dominate the scattering size would become longer than currently estimated. Consequently, the wavelength dependence of the intrinsic size becomes steeper ($\propto \lambda^{1.4 -1.5}$).

The analysis of the closure phases at the highest frequency available in these
observations (86\,GHz) confirmed that the apparent VLBI source structure is
indeed symmetric.  We found no evidence of time variability of the closure
phases from timescales of minutes to days. The lack of any visible structural variation or component ejection after an NIR flare points to an upper velocity limit in the range of of 0.002--0.1\,c.

\begin{acknowledgements}
We thank the anonymous referee for valuable and useful comments.
R.-S. Lu and D. Kunneriath are members of the International Max
Planck Research School (IMPRS) for Astronomy and Astrophysics at the
MPIfR and the Universities of Bonn and Cologne. R.-S. Lu thanks
Z.-Q. Shen for carefully reading an early version of the manuscript and useful comments. We thank Y. Y. Kovalev and K. Sokolovsky for assistance during the data analysis. We would also like to thank A. Brunthaler for useful discussion and
valuable comments.
\end{acknowledgements}
\bibliographystyle{aa}
\bibliography{sgra}

\appendix
\section{Amplitude calibration procedure}
\label{appendix}
First, an a-priori amplitude calibration was performed
using measurements of the antenna gain and system temperature for each telescope. The system temperatures and station gains were then corrected for atmospheric opacity effects. For each station, the zenith opacity was found to vary slightly over epochs, but not in a systematic manner that could produce the final source flux density variations. That the secondary calibrators showed almost no variations in total flux, further reassures that the amplitude variations seen for Sgr\,A* are real.

To remove effects of pointing or focus errors, which in principle could lead to low amplitudes of individual VLBI scans, particularly at 86\,GHz \citep{2001AJ....121.2610D}, we used the upper envelope of the visibility amplitude at the short uv-spacing (between 20 and 100 Mega-lambda) to determine the total flux density. In defining the upper envelope, we also exclude scans where a high opacity correction was necessary. The flux density scale then was also checked using near in time flux density measurements of a number of extragalactic compact radio sources, which were observed
as secondary calibrator sources. During the initial steps of the iterative mapping and self-calibration process, we applied strong uv-tapering and station weighting, which ensured that the flux density on the shortest baselines was maintained. We then used the closure amplitudes and performed an iterative amplitude and phase self-calibration process.
For the gain solutions we  started with solution intervals of several hours.
Subsequent reduction of the solution intervals down to timescales of minutes
removed sidelobes and decreased the residual noise in the maps until,
at the end of the imaging process, a stable and low-noise CLEAN-image was obtained. The fully self-calibrated visibilities were then used for the final Gaussian model fitting.

To further confirm the reliability of our imaging, we also analyzed the closure amplitudes in a way similar to that described by \citet{2004Sci...304..704B} for some selected epochs. In each case we confirm the results obtained via hybrid phase imaging with additional amplitude self-calibration.

\end{document}